\documentclass[iop]{emulateapj}

\usepackage{graphicx}
\usepackage{mathtools}
\usepackage{enumerate}
\usepackage{color}
\newcommand{\kms}{$\,$km$\,$s$^{-1}$}
\newcommand{\kmsmpc}{$\,$km$\,$s$^{-1}\,$Mpc$^{-1}$}
\newcommand{\nnylee}[1]{#1}
\newcommand{\nylee}[1]{#1}
\newcommand{\ylee}[1]{#1}
\newcommand{\magi}{ mag}
\newcommand{\NMpct}{ $N\,$Mpc$^{-2}$}
\accepted{25/02/2016}

\begin{document}

\title{Galaxy Luminosity Function of Dynamically Young Abell 119 Cluster: Probing the Cluster Assembly}
\shorttitle{Luminosity Function of Abell 119}

\author{Youngdae Lee\altaffilmark{1,2}, Soo-Chang Rey\altaffilmark{1,\ylee{6}}, Michael Hilker\altaffilmark{3}, Yun-Kyeong Sheen\altaffilmark{\ylee{2},4}, and Sukyoung K. Yi\altaffilmark{5}}
\shortauthors{Lee et al.}

\altaffiltext{1}{Department of Astronomy and Space Science, Chungnam National University, Daejeon 305-764, Republic of Korea}
\altaffiltext{2}{Korea Astronomy and Space Science Institute, Daejeon 305-348, Republic of Korea}
\altaffiltext{3}{European Southern Observatory, Karl-Schwarzschild-Str.~2, D-85748 Garching, Germany}
\altaffiltext{4}{Departamento de Astronom\'{i}a, Universidad de Concepci\'{o}n,Casilla 160-C, Concepci\'{o}n, Chile}
\altaffiltext{5}{Department of Astronomy and Yonsei University Observatory, Yonsei University, Seoul 120-749, Republic of Korea}
\altaffiltext{6}{\ylee{Author to whom any correspondence should be addressed} (screy@cnu.ac.kr).}

\begin{abstract}

We present the galaxy luminosity function (LF) of the Abell 119 cluster down to $M_r\sim-14${\magi} based on deep images in the $u$-, $g$-, and $r$-bands taken by using MOSAIC II CCD mounted on the Blanco 4m telescope at the CTIO. The cluster membership was accurately determined based on the radial velocity information as well as on the color-magnitude relation for bright galaxies and the scaling relation for faint galaxies. The overall LF exhibits a bimodal behavior with a distinct dip at $r\sim18.5${\magi} ($M_r\sim-17.8${\magi}), which is more appropriately described by a two-component function. The shape of the LF strongly depends on the cluster-centric distance and on the local galaxy density. The LF of galaxies in the outer, low-density region exhibits a steeper slope and more prominent dip compared with that of counterparts in the inner, high-density region. We found evidence for a substructure in the projected galaxy distribution in which several overdense regions in the Abell 119 cluster appear to be closely associated with the surrounding, possible filamentary structure. The combined LF of the overdense regions exhibits a two-component function with a distinct dip, while the LF of the central region is well described by a single Schechter function. We suggest that, in the context of the hierarchical cluster formation scenario, the observed overdense regions are the relics of galaxy groups, retaining their two-component LFs with a dip, which acquired their shapes through galaxy merging process in group environments, before they fall into a cluster.

\end{abstract}


\section{Introduction}

According to the hierarchical structure formation models in the $\Lambda$ cold dark matter cosmology, clusters of galaxies have been built-up as a result of infall of galaxies and galaxy groups along filaments into a cluster and their subsequent merging \citep{Wes95,Fal05,Col05,Ber09}. Once the galaxies are confined to a cluster, they are continuously affected by various environmental processes, such as the ram pressure, harassment, and tidal interaction \citep{Byr90,Par09,Moo96,Moo98,Mos00,Gne03,Bos06,Bos08}. On the other hand, the galaxies in a group may have already evolved before the group is assembled into a cluster \citep[i.e., preprocessing;][]{Zab98,Fuj04,McG09,DeL12,Dre13}. An investigation of the properties and evolution of galaxies in the smaller, local-scale structures of galaxy groups can provide insights into the formation and aggregation of the larger, global-scale structures of galaxy clusters.

For understanding the formation history and evolution of the galaxies in galaxy clusters, the galaxy luminosity function (LF) is a powerful tool for describing the fundamental properties of galaxies. In the early studies, a single Schechter function was suggested as an analytic form to describe the ``universal" LF of \ylee{galaxies in clusters \citep{Sch76} and field environment \citep{Mad02,Bla05}}. However, by using deep and accurate photometry of galaxies, it has been revealed that the LFs of galaxies in many clusters possess more complex features, such as dips at intermediate luminosities and excess of faint galaxies, making a two-component LF a better fit to the observed data \citep[e.g.,][]{Pop06,Mer06,Ban10,Liu11,Yam12}. Moreover, it has been suggested that the LF shape varies with the environment even within a single cluster, providing information on the formation and subsequent evolution of galaxies depending on the cluster environmental effects \citep{Mer06,Pop06}.

Because variations in the galaxy LF are related to the cluster evolutionary state, the LFs of dynamically young and unrelaxed clusters are the best indicators for studying the ongoing assembly of galaxy clusters. According to the hierarchical structure formation scenario, the presence of substructures in a cluster is a signature of cluster formation processes. An indication of LF variations across the cluster substructures was found as the LFs of the substructures exhibit different shapes, depending on their dynamical state \citep[e.g.,][]{Yan04,Zha11}. Except for a few studies, a detailed investigation on the LFs of dynamically young clusters has not received proper attention.

In this study, we chose the galaxy cluster Abell 119 to further investigate the galaxy LFs in young clusters, by examining in particular the LF variation as a function of different local densities and regions in the cluster. The Abell 119 cluster is a nearby ($z\sim0.044$), rich cluster of galaxies with a velocity dispersion of $\sigma_{v}\sim800$\kms{} \citep{Mel81,Fad96,Kat96,Val11}. It has been suggested that the Abell 119 cluster is a dynamically complex cluster with several substructures in the galaxy distribution and the X-ray surface brightness profile \citep{Fab93,Kri97,Ram07,Tia12}. The X-ray data reveal interesting features, such as the positional offset of 36 arcsec between the X-ray and optical peaks, X-ray distribution elongated in the North-East direction, and no indication of an inward cooling flow \citep{Fab93,Per98,Ros10}. In addition, radio observations detected two radio galaxies with narrow-angle tails, which might be associated with recent or ongoing merger activities between the cluster and sub-clusters \citep{Bli98,Fer99}. Consequently, many observational results suggest that the Abell 119 cluster is an example for a dynamically complex, young system \citep{Edg90,Vik09,Ros10}.

In this paper, we investigate the behavior of the LF of the Abell 119 cluster, based on a statistically significant sample of member galaxies, constructed from MOSAIC II CCD observations by using the Blanco 4m telescope at the CTIO. This paper is structured as follows. In Section 2, we describe the observations and the data reduction process. To construct an accurate LF, we performed the detection completeness\nylee{. We also selected cluster member galaxies based on the radial velocity information as well as on the color-magnitude relation for bright galaxies and the scaling relation for faint galaxies.} In Section 3, we analyze the overall cluster LF and the environmental dependence of the LFs of galaxies within the Abell 119 cluster. The large-scale spatial distribution of galaxies in the cluster and the LFs of overdense regions are also explored. Finally, Section 4 contains the discussion \nylee{on the bimodal LFs of dynamically young clusters constraining the assembly history of galaxy clusters}.

%we investigate the behavior of the LF of the Abell 119 cluster, based on a statistically significant sample of member galaxies, constructed from MOSAIC II CCD observations by using the Blanco 4m telescope at the CTIO. 

\section{Data and Analysis}
\subsection{Data}

Our investigation makes use of the imaging and the integrated photometric data of \citet{She12}. As a part of the survey of galaxies in the nearby, rich Abell clusters, $u$-, $g$-, and $r$-band deep images of galaxies in the Abell 119 cluster were taken by using the MOSAIC II CCD that was mounted on the Blanco 4m telescope at the CTIO (see \citealt{She12} for details). The overall dimension of the MOSAIC II CCD is 8K $\times$ 8K pixels with the pixel scale of 0.24$''$ pixel$^{-1}$, which corresponds to the 36 $\times$ 36 arcmin$^2$ field of view. \ylee{The images are based on a single MOSAIC II pointing combined from more than five dither positions in each filter (see Table 1 of \citet{She12}).} The total exposure times of the images were 6000, 5760, and 5760 seconds for the $u$-, $g$-, and $r$-band, respectively. The average seeing was under 1 arcsec. After performing basic pre-processings and flux calibration with standard stars in CDF-S \citep{Smi03}, the optical integrated magnitudes of the galaxies were measured \ylee{in the AB magnitude system} using SExtractor \citep{Ber96}. Objects with CLASS\_STAR $\leq$ 0.8 from our sample were selected as galaxies. Galactic foreground extinction was corrected using the reddening maps from \citet{Sch98}. See \citet{She12} for more details on the observations, data reduction, and integrated photometry. For the rest of the analysis in this paper we used the distance modulus of 36.30, which is obtained from our radial velocity distribution of galaxies by assuming $H_0 = 73$\kmsmpc{} (see Section 2.3 for details). We also adopted the location of UGC\,579 (R.A. (J2000) $=$ 00$^h$ 56$^m$ 16.1$^s$, Decl. (J2000) $=$ -01$^\circ$ 15$'$ 19.1$''$) as the cluster center of Abell 119.

\subsection{Surface Photometry and Photometric Completeness} 

For the surface photometry of our sample galaxies, we cropped the $r$-band image around each galaxy. All of the objects detected by the SExtractor, except the galaxy itself, were masked using the segmentation image, following which the SExtractor was performed again on the masked image. The background subtracted image was constructed using the background map returned from the SExtractor. We determined the surface brightness profiles of our galaxies using the ELLIPSE task in the IRAF. We fixed the center of the isophote to the one measured by the SExtractor and set the position angle and ellipticity as the free parameters.

\begin{figure}
\epsscale{1.0}
\plotone{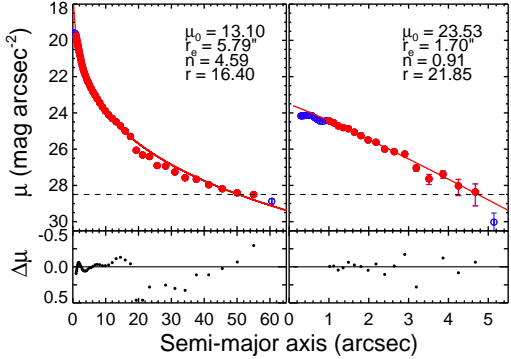}
\caption{
Examples of S\'ersic fits to the surface brightness profiles of a bright (left panel) and a faint (right panel) galaxy. Red filled circles indicate the data used for the S\'ersic fit. Blue open circles are the data points excluded from the fitting; these points are located in the inner ($<$1 arcsec) region or below the $+1\sigma$ deviation of the sky background level (horizontal dashed lines). Error bars denote the root mean square (rms) deviation of the surface brightness for each isophote. In each panel, the red solid curve is the best-fit S\'ersic model. The parameters of the best-fit S\'ersic model and the apparent galaxy magnitude are shown in the top right corner of each panel. The bottom panels show the residuals between the best-fit model profiles and the observed surface brightness profiles.
}
\label{SurfEx}
\end{figure}

We fitted the radial surface brightness profiles by the S\'ersic functions, using the following equation \citep{Gra05}. 
\begin{equation}
\mu(R)=\mu_0 + \frac{2.5}{\ln(10)} \left(\frac{R}{h}\right)^{1/n},
\end{equation}
where $\mu_0$ is the central surface brightness and $n$ is the S\'ersic index. The parameter $h$ is the scale-length that is related to the effective radius ($R_e$) as $R_e = b^nh$, with $b = 1.9992n - 0.3271$ for $0.5 < n < 10$ \citep{Gra05}. The fitting was performed using the IDL routines in the MPFIT package \citep{Mar09}. To minimize the seeing effects, the inner ($<$1 arcsec) region was excluded from the fitting. Furthermore, the data points below the $+1\sigma$ deviation of the background level were excluded as well. In Figure~\ref{SurfEx}, we present examples of the S\'ersic profile fitting for a bright (left panel) and a faint (right panel) galaxy, where the solid curves are the best-fit S\'ersic models.

\begin{figure}
\epsscale{0.8}
\plotone{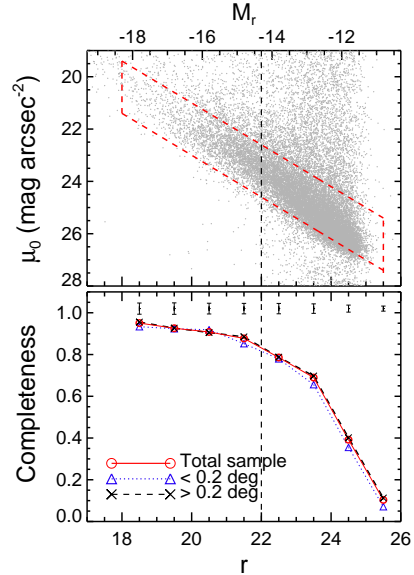}
\caption{
(Top) The $r$-band central surface brightness ($\mu_0$) of the observed objects in the Abell 119 cluster as a function of \ylee{$r$}-band magnitude. Gray dots are the extended sources within the observed field. The red dashed box indicates the range of input parameters for constructing the artificial galaxies. (Bottom) Completeness curves as a function of magnitude. The red solid line denotes the results for the entire sample. The blue dotted and black dashed lines are for the inner ($<$0.2 deg) and outer ($>$0.2 deg) regions of the cluster, respectively. For each magnitude bin, the error bars correspond to the Poisson error. The black vertical dashed line indicates our magnitude limit ($r = 22${\magi}) for completeness above $\sim$80\%.
}
\label{Completeness}
\end{figure}

To study the galaxy LF, completeness corrections have to be applied to the observed number counts of galaxies. To estimate the completeness, we generated 10000 artificial galaxies by adopting the structural parameters obtained from the surface photometry of the observed galaxies. Figure~\ref{Completeness} (top panel) shows the $r$-band central surface brightness ($\mu_0$) and the $r$-band magnitude ($r$) of the objects that were observed in the Abell 119 cluster (gray dots). The $\mu_0$ was obtained from the best-fit S\'ersic model. To construct the artificial galaxies, first, $\mu_0$ and $r$ were randomly selected following the parameter ranges of the observed galaxies (see red dashed box in the top panel of Fig. ~\ref{Completeness}). The S\'ersic indices ($n$) were also randomly selected in the range of $0.5 < n < 4$ \citep{Bal07,Gra08}. The effective radii of the artificial galaxies were calculated with the selected $\mu_0$, $r$, and $n$, using the equations of \citet{Gra05}. Two hundred fifty out of the 10000 artificial galaxies were randomly distributed in the observed image and then were searched and measured using the SExtractor with the same configuration parameters as those that were used in our original photometry. This procedure was repeated for 40 times.

We calculated the detection rate of the artificial galaxies as a function of magnitude (the bottom panel of Fig.~\ref{Completeness}). In general, the completeness decreases with the magnitude (red solid line for the total sample). More than 80\% of the simulated galaxies at $r < 22${\magi} were recovered by the SExtractor. However, there is significant incompleteness regarding faint galaxies dropping below 50\% for $r > 24${\magi}. A completeness correction for the faint galaxies might result in an inaccurate LF slope. Therefore, we restricted the sample to the galaxies with $r < 22${\magi} for the analysis of LFs in the Abell 119 cluster. To examine the spatial variation of the completeness, we also calculated the detection rate for the inner ($<$0.2 deg, blue dotted line) and outer ($>$0.2 deg, black dashed line) regions of the cluster. While the completeness of the inner region appears to be slightly lower than that of the outer region at faint magnitudes, no significant difference between them was found within the errors ($0.02 \pm 0.03$\% for the mean), implying no severe spatial dependence of the completeness.

\subsection{Membership of galaxies in the Abell 119 cluster}
For an accurate derivation of the LF, it is important to define the cluster membership of galaxies. Although radial velocity information from spectroscopic observations can be directly used to separate the cluster members from the background galaxies, this is generally limited to bright galaxies with high surface brightness. Alternatively, many studies have used a statistical subtraction of the foreground/background galaxies based on the galaxy counts in the nearby or blank regions around the cluster, for eliminating the contamination in the LF \citep{Bar12,Pra05,Pop06}. However, this statistical method is hampered by the density fluctuation around the cluster \citep[e.g.,][]{Hra00}. This effect is likely to be more severe in dynamically young clusters with substructures as well as at the intersections of filaments such as the Abell 119 cluster \citep[][see also Section 3.3]{Fab93,Kri97,Ram07,Tia12}.

Another indirect method for determining the galaxy membership is to use the scaling relations (and morphology) of galaxies \citep{Bin84,Fer88,Fer89,Hil03,Mie07,Chi10,Bos11}. At a given magnitude, the background galaxies with high surface brightness can be separated from the cluster member galaxies with low surface brightness. \citet{Chi10} tested the success rate for determining the membership of the Coma cluster galaxies by using scaling relations, and concluded that this method is reliable even for faint, low surface brightness galaxies. Here we consider the available radial velocity information as well as the color-magnitude relation for bright ($r < 19${\magi}) galaxies in combination with the scaling relation for faint galaxies ($r > 19${\magi}), for the selection of the member galaxies in the Abell 119 cluster.

\subsubsection{Bright Galaxies}

To determine the spectroscopic membership of bright galaxies in the Abell 119 cluster, we have collected the data on the radial velocities of 193 galaxies from the SDSS DR9 \citep{Ahn12}. To complete the data for the galaxies for which radial velocities were not measured in the SDSS, we added additional 187 galaxies from the independent spectroscopic measurements available in the NASA/IPAC Extragalactic Database (NED) and from previous studies of the Abell 119 cluster \citep{Fab93,Kat98,Rin03,Smi04,Cav09}. Overall, we secured 380 galaxies with known radial velocities. These galaxies cover the magnitude range of $13.6${\magi} $ < r < 24.6${\magi}. To test the consistency between the SDSS and other data, we examined the radial velocities of 116 galaxies for which we had two or more independent radial velocity measurements. The mean difference (25\kms{}) between the measured radial velocities was negligible, within the standard deviation (52\kms{}).

\nylee{Our compiled bright galaxies with available spectroscopic radial velocity information are not biased towards a targeted cluster selection. Out of 281 bright ($r < 19${\magi}) galaxies with spectroscopic data, 137 galaxies ($\sim$49\%) are from the SDSS in which selected galaxies are randomly sampled. One hundred thirty-one galaxies ($\sim$47\%) are from \citet{Cav09}. They also selected a random sample with $V < 21.5${\magi} and $B-V < 1.4${\magi} (i.e., $g-r < 1.3${\magi}) to avoid any bias in the observed morphological type \citep[See section 2 of ][]{Cav09}. Therefore, the majority ($\sim$96\%) of our bright sample galaxies with spectroscopic data are randomly selected ones without any systematic bias.}

\begin{figure}
\epsscale{1.0}
\plotone{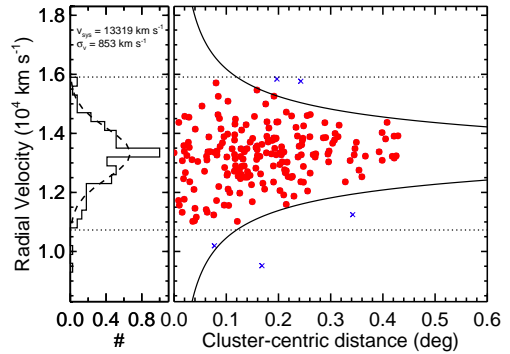}
\caption{
(Left) The radial velocity distribution for the Abell 119 galaxies with available radial velocities (see histogram). The black dashed line is the best-fit Gaussian function to the sample. The determined systemic velocity ($v_{\rm sys}$) and velocity dispersion ($\sigma_v$) of the Abell 119 cluster are also indicated in the plot. (Right) The radial velocities of the galaxies as a function of the cluster-centric distance. The selected spectroscopic member and background galaxies are marked by the red filled circles and blue crosses, respectively. \ylee{The background (178) and forground (4) galaxies with large or small radial velocities were not captured by this plot.} The black solid lines are the caustic curves of the infall model, and the horizontal dotted lines delineate the region of $\pm3\sigma_v$ away from the $v_{\rm sys}$.
}
\label{VelInfo}
\end{figure}

\begin{figure}
\epsscale{1.0}
\plotone{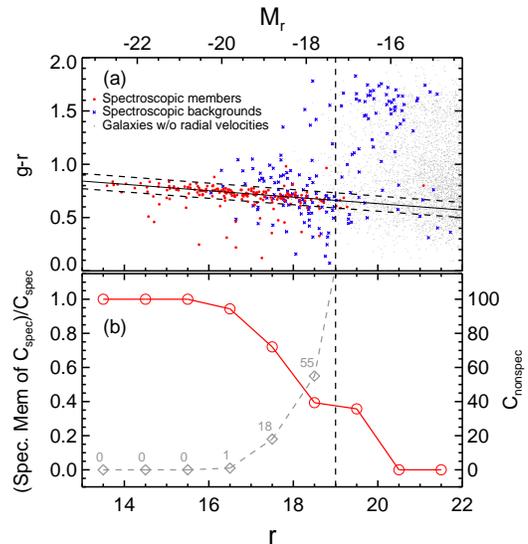}
\caption{
(a) The color-magnitude diagram of galaxies in the Abell 119 cluster. The red filled circles and the blue crosses correspond to the spectroscopically selected member (M$_{spec}$) and background (B$_{spec}$) galaxies, respectively. The gray dots are the galaxies for which no radial velocity information is available. The black solid line is the red sequence determined by the linear least square fit to the sample of M$_{spec}$. The black dashed lines are the $\pm3\sigma$ deviations from the red sequence. (b) The fraction of the number of spectroscopic members in the C$_{spec}$ sample to the number of all C$_{spec}$ galaxies (red solid curve), where C$_{spec}$ is the number of galaxies with radial velocities that are bluer than the upper envelope of the red sequence (i.e., C$_{phot}$ with radial velocities). The gray dashed curve is for the trend of C$_{nonspec}$, which are C$_{phot}$ galaxies without radial velocities. The actual number of C$_{nonspec}$ galaxies for each magnitude bin is also indicated. Spectroscopic incompleteness is corrected only for galaxies with $r < 19${\magi} (see the vertical dashed line).
}
\label{Membership}
\end{figure}

The membership assignment of bright galaxies with radial velocities was accomplished in several steps. First, the distribution of radial velocities was fitted with a Gaussian function (black dashed curve in the left panel of Figure~\ref{VelInfo}), following which the systemic velocity ($v_{sys}$) and the velocity dispersion ($\sigma_v$) of the cluster were determined. The virial radius ($r_{200}$) of the cluster was then calculated using the equation given by \citet{Car97}:
\begin{equation}
r_{200} = \frac{\sqrt{3}\sigma_{v}}{10H_0},
\end{equation}
where the $H_0$ is 73\kmsmpc{}. Second, we examined the radial velocity distribution as a function of the cluster-centric distance (right panel of Fig.~\ref{VelInfo}). In this distribution, we overplotted a spherical symmetric infall model (solid curves in the right panel of Fig.~\ref{VelInfo}) using the measured $\sigma_v$ and $r_{200}$ \citep{Pra94}. The infall model encompasses all galaxies in a cluster whose infall motion is decoupled from the Hubble flow. Only galaxies that were bounded by the infall model lines were selected. Furthermore, the galaxies with radial velocities larger than 3$\sigma_v$ away from the $v_{\rm sys}$ (dotted lines in Fig.~\ref{VelInfo}) were also rejected and the process was repeated iteratively until ensuring convergence. Finally, 193 and 187 out of 380 galaxies were defined as spectroscopic members (M$_{\rm spec}$) and background galaxies (B$_{\rm spec}$), respectively. Most spectroscopic member galaxies cover a relatively bright magnitude range of $13.6${\magi} $ < r < 19.3${\magi} (see also Figure~\ref{Membership}a). Using the selected sample of M$_{\rm spec}$ , the final values of $v_{\rm sys}$, $\sigma_v$, and $r_{200}$ were determined as 13$\,$319\kms{}, 853\kms{}, and 2.0 Mpc (i.e., 0.63 deg), respectively. The derived $v_{\rm sys}$ and $\sigma_v$ are in good agreement with the previously obtained values \citep[e.g.,][]{Fab93,Tia12}. However, our $\sigma_v$ is different from that (618\kms{}) reported by \citet{She12} which was measured using a bright galaxy sample with $M_r < -20${\magi}. The corresponding distance and distance modulus of the Abell 119 cluster are 182 Mpc and $m-M = 36.30${\magi}, respectively, consistent with the previously reported values \citep{Mel81,Kat96,Val11,Tia12}.

To further determine the membership of bright galaxies without available radial velocities (e.g., gray dots in Figure~\ref{Membership}a), we relied on the color-magnitude distribution of galaxies (Figure~\ref{Membership}a). The red sequence (black solid line) was determined by the linear least square fitting and by the sigma clipping method applied to the sample of M$_{\rm spec}$ (red filled circles):
\begin{equation}
g-r=-0.030(\pm0.001)r + 1.234 (\pm0.0237).
\end{equation}
First of all, we rejected galaxies that are redward of the red sequence (i.e., larger than three times the average color dispersion of the red sequence, see upper dashed line in Fig.~\ref{Membership}a), because there is a lack of cluster member galaxies significantly redward of the red sequence \citep[e.g.,][]{Gla00,Muz08,Rin08,Lu09,Bar12}. We considered bright galaxies that are bluer than this upper red sequence envelope as possible cluster member candidates (hereafter C$_{phot}$).

We estimated a correction factor that indicates how many C$_{phot}$ galaxies without available radial velocities are true member galaxies \citep{Ada07,Yoo08}. For this, we defined C$_{spec}$ and C$_{nonspec}$ as the galaxies with and without radial velocities respectively, that are within our selection boundaries (i.e., C$_{nonspec}$ $=$ C$_{phot} - $ C$_{spec}$). For a given magnitude, we defined the correction factor as the ratio of the number of spectroscopically confirmed cluster members among the C$_{spec}$ galaxies to the number of all C$_{spec}$ galaxies (red solid curve in Fig.~\ref{Membership}b). This ratio gradually decreases from $r\sim16${\magi} faintward, with most C$_{spec}$ galaxies brighter than this magnitude being the cluster members. Therefore, at each magnitude bin, a correction factor can be applied to the number of C$_{nonspec}$ galaxies (gray dashed curve in Fig.~\ref{Membership}b) to estimate the fraction of possible member galaxies among them. Note that we applied this correction factor only to the bright galaxies with $r < 19${\magi} (vertical dashed line in Fig.~\ref{Membership}), because the spectroscopic coverage of fainter galaxies drops below 40\%, providing less statistical meaning. 

\ylee{Note that there are a few spectroscopic member galaxies with colors redder than the upper red sequence envelope in the color-magnitude diagram (see five galaxies brighter than $r=19${\magi} in Fig. 4(a)). To account for them, we need to estimate an additional statistical correction factor that indicates how many galaxies with colors redder than this envelope are true member galaxies. For this, we considered bright ($r < 19${\magi}) galaxies without available radial velocities, that are redder than the upper red sequence envelope  (hereafter C$_{nonspec,red}$). We defined the correction factor as the ratio of M$_{spec,red}$/C$_{spec,red}$, where M$_{spec,red}$ and C$_{spec,red}$ are spectroscopic member galaxies and galaxies with radial velocities, respectively, that are both above the upper red sequence envelope. For a given magnitude bin, this correction factor was applied to the number of  C$_{nonspec,red}$ galaxies for the number of possible member galaxies among them. Finally, we obtained 0, 0.6, and 3.1 possible member galaxies for the magnitude bins of  $r < 17${\magi}, 17{\magi} $ < r < $ 18{\magi}, and 18{\magi} $ < r < 19${\magi}, respectively.  To construct the LF for the bright galaxies, the corrections for C$_{nonspec}$ and C$_{nonspec,red}$ were added to the number M$_{spec}$ of member galaxies for each magnitude bin.}

\subsubsection{Faint Galaxies}

The majority of low-luminosity galaxies with low surface brightness exhibit exponential radial profiles and follow a well-known correlation between the central surface brightness and magnitude \citep{Bin84,Fer88,Fer89}. This scaling relation allows to discriminate faint cluster members from background galaxies when the no radial velocity information is available \citep{Kar03,Hil03,Mie07,Rin08,Chi10,Bos11}. For a given magnitude, background galaxies have higher central surface brightness than the cluster members. In previous studies, this method of membership determination for faint galaxies was confirmed as reliable, compared with the direct spectroscopic method \citep[e.g.,][]{Rin08,Chi10}.

\begin{figure}
\epsscale{1.0}
\plotone{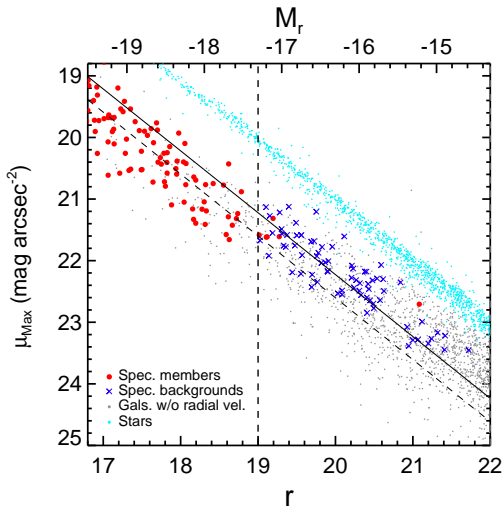}
\caption{
The peak surface brightness ($\mu_{Max}$) vs. the apparent $r$-band magnitude of the galaxies in the Abell 119 cluster. The red filled circles are the data for the spectroscopically confirmed member galaxies. The blue crosses are the data for the spectroscopically confirmed background galaxies with $r > 19${\magi}. The gray dots are the galaxies that are more than $+3\sigma$ blueward from the red sequence in the color-magnitude relation. The thin dashed line is the least square fit to the spectroscopically confirmed member galaxies in the range of $17${\magi} $< r < 19${\magi}. The black solid line represents the segregation between the members and the background galaxies for our faint sample with $r > 19${\magi}. Note that the stars (cyan dots) defined by CLASS\_STAR $> 0.8$ exhibit a tight sequence that is well separated from the galaxies.
}
\label{Membership2}
\end{figure}

In Figure~\ref{Membership2}, we show the peak surface brightness ($\mu_{Max}$) obtained from the SExtractor vs. the $r$-band magnitude of galaxies in the Abell 119 cluster. The $\mu_{Max}$ is defined as the surface brightness of the brightest pixel in the center of a galaxy \citep{Ber96}\nylee{, which is comparable to the central surface brightness.} For reducing the contamination of our sample, we only considered those galaxies that are blueward of the $+3\sigma$ deviation from the red sequence in the color-magnitude relation (gray dots in Fig.~\ref{Membership2}). \nylee{The shape of the scaling relation strongly depends on the definition of the adopted surface brightness parameter \citep{Gra03}. In the case of using the central surface brightness, the scaling relation presents a linear relation \citep[see Fig. 12 of ][]{Gra03}. Therefore,} with spectroscopically confirmed members in the range of 17{\magi} $< r < 19${\magi} \ylee{($-19.3${\magi} $< M_r < -17.3${\magi}),} a robust least square fitting was performed in the plane of $\mu_{Max}$ and $r$ (thin dashed line in Fig.~\ref{Membership2}). The $+1\sigma$ deviation from this relation defines our upper boundary to properly cover most spectroscopic member galaxies (black solid line in Fig.~\ref{Membership2}). This sequence separates the possible members (occupying the region below the black solid line in Fig.~\ref{Membership2}) from the background galaxies (above the black solid line in Fig.~\ref{Membership2}) in the faint galaxy sample with $r > 19${\magi}. 

\ylee{Distinguishing between member galaxies and background objects becomes somewhat difficult at faint magnitudes, because the surface brightness measurements are mainly affected by the spatial resolution of the data \citep{Kar03}. In our case, a fraction (30\%, 22 of 74) of spectroscopically confirmed background galaxies (blue crosses in Fig. 5) in the range of 19{\magi} $ < r < 22${\magi} are classified as member galaxies (i.e., below the black solid line). For the statistical correction of the background galaxy contamination at 19{\magi} $ < r < 22${\magi}, we calculated the correction factor at each magnitude bin:
\begin{equation}
F_{cor} = 1 -  \frac{B_L}{B_U}\frac{P_U}{P_L},
\end{equation}
where B$_U$ and B$_L$ are the number of spectroscopic background galaxies above and below the black solid line of Fig. 5, respectively. The P$_U$ and P$_L$ are the numbers of galaxies without radial velocity information (i,e., gray dots in Fig. 5) above and below the black solid line, respectively. The correction factors are 0.77, 0.81, and 1.00 for the magnitude bins of 19{\magi} $ < r < 20${\magi}, 20{\magi} $ < r < 21${\magi}, and 21{\magi} $ < r < 22${\magi}, respectively. This indicates that 23\%, 19\%, and 0\% of member galaxies defined from the scaling relation are contaminated by background galaxies at 19{\magi} $ < r < 20${\magi}, 20{\magi} $ < r < 21${\magi}, and 21{\magi} $ < r < 22${\magi}, respectively. To construct the LF for the faint galaxies, the P$_L$ was multiplied by the correction factor for each magnitude bin.
}

\ylee{We also examined the contamination of foreground galaxies using bright galaxies with available radial velocity information. We considered seven bright foreground galaxies, which are below the lower boundary of the caustic curve or $-3\sigma$ away from the systemic velocity of the Abell 119 (see Fig. 3). We defined the foreground galaxy fraction as the ratio of the number of foreground galaxies to the number of foreground and spectroscopic member galaxies (i.e., 200 galaxies). Assuming that the fraction of foreground galaxies is similar in all luminosity bins, we suggest that the foreground galaxy contamination to the faint member galaxy sample is negligible (i.e., about 3.5\%).}

\section{Results}
\subsection{Galaxy Luminosity Function}
\begin{figure}
\epsscale{0.7}
\plotone{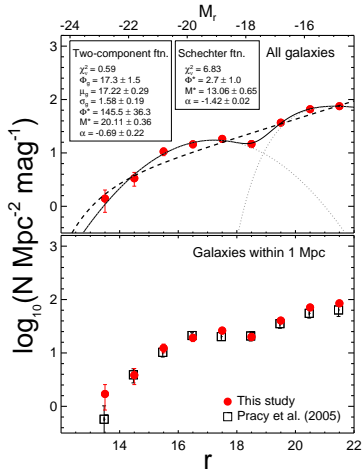}
\caption{
(Top) The $r$-band luminosity function of all galaxies in the Abell 119 cluster. The red filled circles are completeness corrected galaxy counts from this study. The count errors for all magnitude bins represent Poisson statistics. The dashed curve is the fit with a single Schechter function. The solid curve is the fit with the two-component LF (i.e., the sum of Gaussian and Schechter functions), and the dotted curves indicate its individual components. The best-fit parameters of the single Schechter function and the two-component function are shown as well. For all best-fit parameters, the errors were estimated from the covariance matrix. (Bottom) Comparison of our LF (red filled circles) within 1 Mpc radius from the cluster center with that obtained by \citet{Pra05} (open squares).
}
\label{LumF}
\end{figure}

\ylee{Based on 916 galaxies (257 bright and 659 faint galaxies) and the detection completeness as well as membership corrections  described in the previous section, we constructed the galaxy LF for the $r$-band in the range of 13{\magi} $ < r < 22${\magi} ($-23.3${\magi} $ < M_r < -14.3${\magi}), as shown in Figure 6.} We first fitted the conventional single Schechter function to the LF of all galaxies (dashed curve in the top panel of Fig.~\ref{LumF}):
\begin{equation}
\Phi(M) = \Phi^{*}10^{-0.4(M-M^{*})(\alpha+1)} e^{-10^{-0.4(M-M^{*})}},
\end{equation}
where $\Phi^*$ is the characteristic number density, $M^*$ is the characteristic absolute magnitude, and $\alpha$ is the faint-end slope of the LF. The $M^*$ may be taken as a measure of the mean luminosity of giant galaxies, and the slope α measures the relative abundance of faint galaxies. The observed LF was only moderately represented by the Schechter function, as revealed by a relatively large reduced chi-square value ($\chi^2_{\nu}$), and the details of the LF (in particular an apparent dip at $r\sim18.5${\magi}) were not captured.

On the other hand, a two-component fit (i.e., a sum of Gaussian and Schechter functions) has been commonly used for describing the observed LF in many clusters \citep{Dri94,Sec96,Tre02,Hil03,Mah05,Pop06,Ban10,Yam12}. This fitting much better describes the LFs of bright and faint galaxy populations. Following \citet{Tre02}, we fitted our data with a two-component LF:
\begin{equation}
\begin{multlined}
\Phi(M) = \Phi_g e^{-(M-\mu_{g})^2/(2{\sigma_g}^2)} \\
 + \Phi^{*}10^{-0.4(M-M^{*})(\alpha+1)} e^{-10^{-0.4(M-M^{*})}}.
\end{multlined}
\end{equation}
The contribution of bright giant galaxies is described by the first term of the Gaussian function, which is characterized by normalization ($\Phi_g$), peak magnitude ($\mu_g$), and dispersion ($\sigma_g$). The second term is the Schechter function that describes the faint dwarf galaxies. The $M^*$ is the characteristic magnitude describing the fractional population of faint galaxies. The $\Phi^*$ and $\alpha$ denote the normalization and the faint-end slope of the fitted LF, respectively.

It is clear that the sum of Gaussian and Schechter functions better fits our observed galaxy counts (solid curve in the top panel of Fig.~\ref{LumF}). One interesting feature is the distinct dip at $r\sim18.5${\magi} ($M_r\sim-17.8${\magi}), which distinguishes the bright galaxies from the faint ones. The presence of a dip has been discussed in many studies, and the dips found in other clusters are present at comparable absolute magnitudes \citep{Hun00,Gon06,Men06,Pop06,Rin08,Agu14}.

\citet{Pra05} reported the LF of galaxies in the Abell 119 cluster using an extensive sample based on deep, wide-field $V$-band mosaic imaging observations. They applied a statistical correction to account for the contamination by the field galaxy population. While their observations cover an overall projected area of 0.85 deg$^2$, corresponding to 9.7 Mpc$^2$ for $H_0 = 70$\kmsmpc{}, they only provided the central LF within 1 Mpc from the cluster center (see their table 2), which is comparable to the areal coverage (0.36 deg$^2$ and 3.6 Mpc$^2$ for $H_0 = 73$\kmsmpc{}) of our observations. In the bottom panel of Fig.~\ref{LumF}, we compare our LF (red filled circles) within 1 Mpc from the cluster center with that reported by \citet{Pra05} (open squares). Their $V$-band LF was converted into an $r$-band LF through proper color conversion using the equation of \citet{Lup05}\footnote{\url{http://www.sdss3.org/dr9/algorithms/sdssUBVRITransform.php}}. The absolute magnitude was estimated by adopting the distance modulus (36.53{\magi}) used by \citet{Pra05}. The general shape of the LF of \citet{Pra05} is quite consistent with ours, including the dip at $r\sim18.5${\magi}.

\subsection{The Environmental Dependence of the LF}
\begin{figure*}
\epsscale{1.0}
\plotone{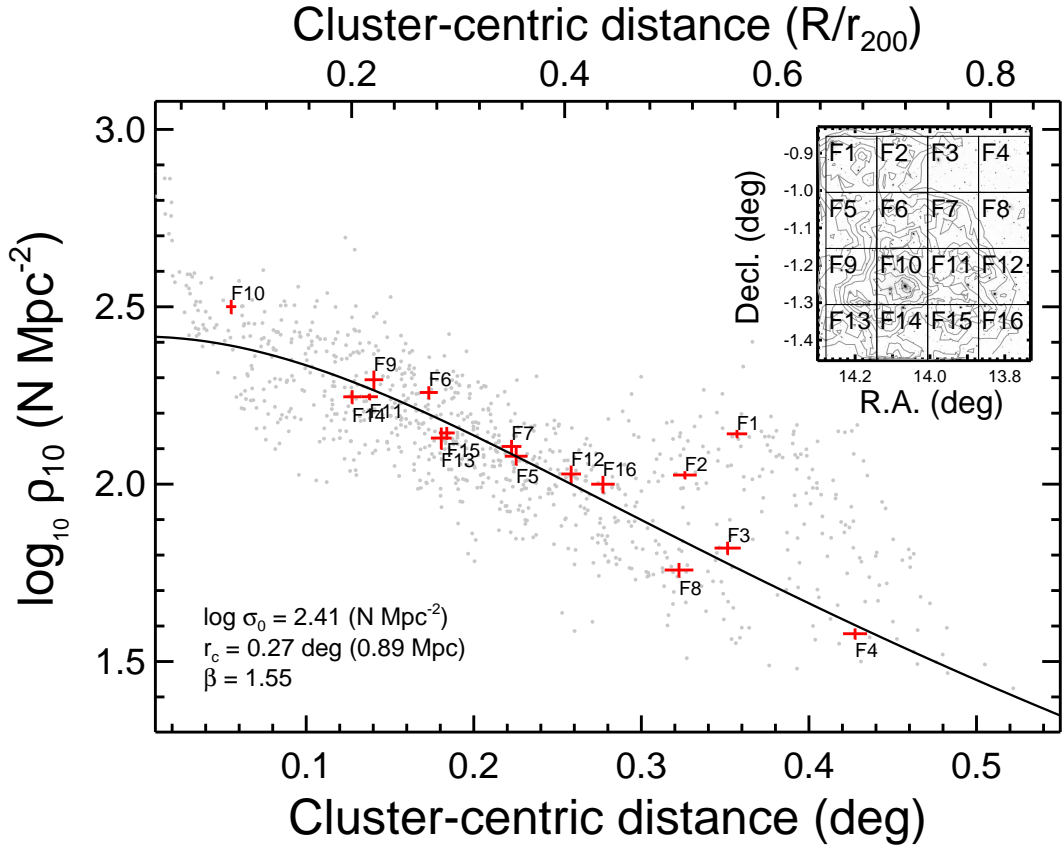}
\caption{
The distribution of the projected local density (log$_{_{10}}\rho_{10}$) as a function of the cluster- centric distance. The solid curve is the best-fit King model profile. The best-fit parameters of the King model are shown as well. The red crosses are the biweight mean local densities and the cluster-centric distances of 16 sub-regions of the Abell 119 cluster, and their sizes correspond to the bootstrap errors. \ylee{The cluster-centric distance normalized by the viral radius is shown in the upper X-axis.} The inset shows the partitioning into 16 subregions and the contours of the distribution of the local projected galaxy density.
}
\label{RadialDensity}
\end{figure*}

It has been known that the LFs of galaxies exhibit different shapes and slopes in different environments within the clusters \citep{Pro03,Pop06,Ada07}. Before investigating the environmental dependence of the LF in the Abell 119 cluster, we examined the correlation between the cluster-centric distance as a measure of the global cluster environment and the projected local density ($\rho_{10}$), derived using the 10 nearest neighbor galaxies with $M_r < -16${\magi}, as a proxy of local environment (see Figure~\ref{RadialDensity}). As expected, $\rho_{10}$ decreases with increasing the cluster-centric distance. We fitted the distribution with a King model profile as defined by \citet{Ada98}:
\begin{equation}
\sigma(r)=\sigma_0\left(\frac{1}{1+\left(\frac{r}{r_c}\right)^2}\right)^{\beta}+\sigma_b,
\end{equation}
where $\sigma_0$, $r_c$, $\beta$, and $\sigma_b$ are the central number density, the core radius, the slope, and the background density of the model profile, respectively. Because only the cluster members are considered, we set $\sigma_b=0${\NMpct}. In order to minimize possible contamination due to the existing sub-clusters on the outskirts of the Abell 119 cluster \citep[e.g.,][]{Tia12}, only the galaxies bounded between 0.05 \ylee{deg} and 0.3 deg from the cluster center are used for the King model fit. As suggested by \citet{Tia12}, it is evident that some galaxies deviate from the model profile, with a significant density peak for the cluster-centric distances in the range of 0.3-0.45 deg. The dominating feature, however, is that the overall galaxy distribution satisfactorily follows the King model profile (solid curve in Fig.~\ref{RadialDensity}). By dividing our observed field into 16 sub-regions with the same area of 73.4 arcmin$^2$ (see inset of Fig.~\ref{RadialDensity}), we identified eight overdense regions (F1, F2, F3, F6, F7, F9, F12, and F16), in which the local galaxy densities, considering the bootstrap errors, are above the best-fit King model profile for a given cluster-centric distance. The region F10 corresponds to the densest central region of the Abell 119 cluster, including the peak in the galaxy density contour map and the central cluster galaxy UGC\,579.

\begin{figure*}
\epsscale{1.0}
\plotone{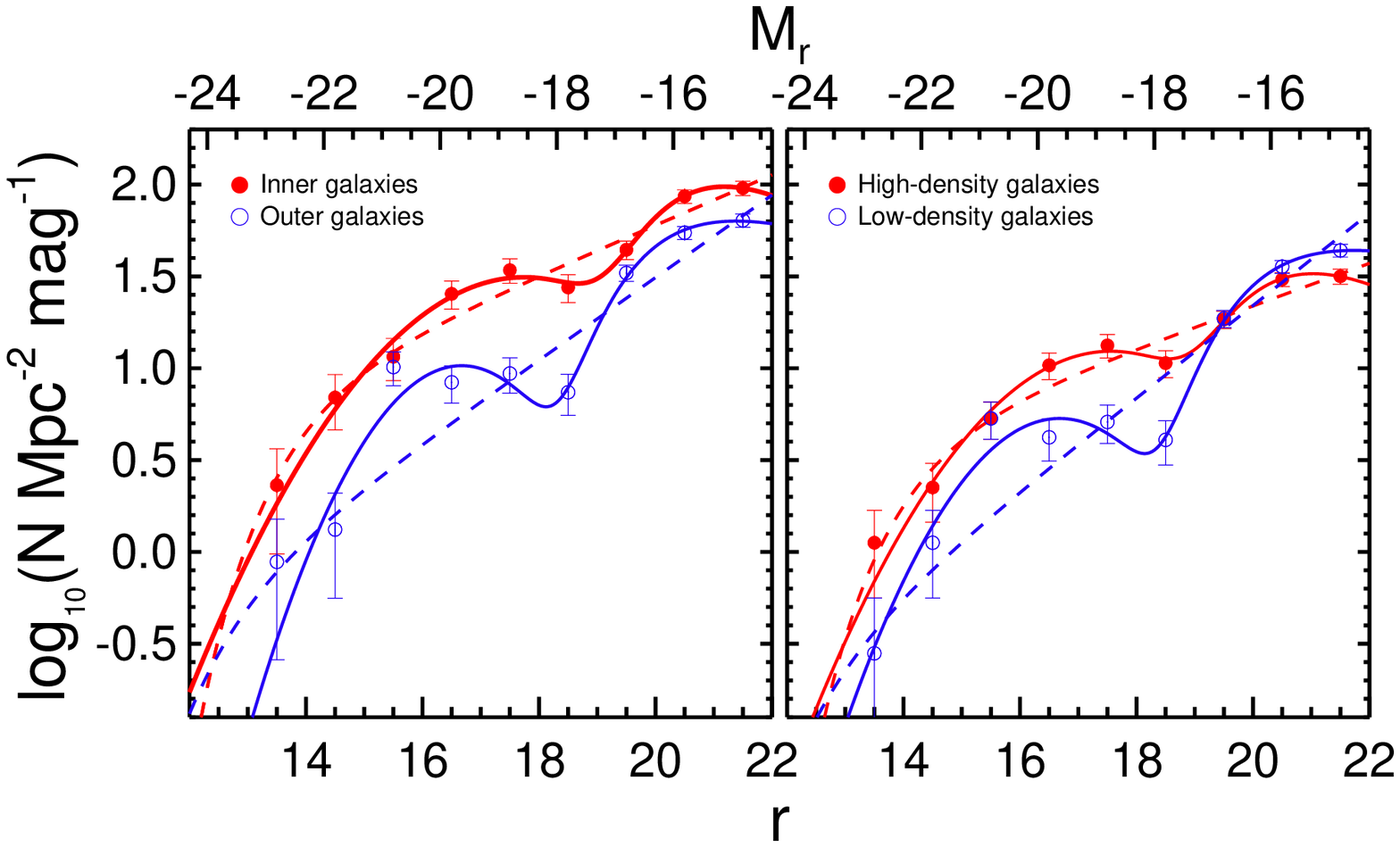}
\caption{
The LFs of galaxies in the Abell 119 cluster, as a function of the environment: inner vs. outer regions (left) and high-density vs. low-density regions (right). The LFs for the inner ($<$0.20 deg from the cluster center) and high-density (log$_{_{10}}\rho_{10} > 2.1${\NMpct}) regions are shown as the red filled circles, and the LFs for the outer ($>$0.20 deg from the cluster center) and low-density (log$_{_{10}}\rho_{10} < 2.1${\NMpct}) regions are shown as the blue open circles. The solid and dashed curves represent the two- component and single Schechter function fits, respectively. The error bars for the number counts are Poisson statistics.
}
\label{EnvLF}
\end{figure*}

In Figure~\ref{EnvLF}, we show the LFs of galaxies for different environments within the Abell 119 cluster. The inner and the outer regions of the Abell 119 cluster are defined as those inside and outside of the 0.20 deg of cluster-centric distance, respectively, which delimits the area of intracluster medium emitting X-ray emission (see the large dotted circle and the blue contours in the bottom panel of Figure~\ref{SpacialGal}). We also considered high- and low-density regions, divided by the local density (log$_{_{10}}\rho_{10}=2.1${\NMpct}) which corresponds to the local galaxy density at 0.2 deg of cluster-centric distance in the King model profile in Fig.~\ref{RadialDensity}. In all panels, the red filled circles and the blue open circles are the data for the galaxies in the inner/high-density and outer/low-density regions, respectively.

We attempted to fit the LFs in the different environments using a single Schechter function (dashed lines in Fig.~\ref{EnvLF}) which allows to examine the variations in the overall LF shape. Because in all cases the LFs of galaxies are inadequately fitted by the single Schechter function, we focused on the slope values for the power-law part. It is worth noting that the slopes \ylee{($-1.57$/$-1.63$)} for the outer/low-density regions are systematically steeper than those \ylee{($-1.34$/$-1.28$)} for the inner/high-density regions (see Table~\ref{table1}). By contrast, \citet{Pra05} found no significant systematic LF slope variations with the cluster-centric distance in the Abell 119 cluster. We also fitted the LFs with the two-component function: red and blue solid curves are for the inner/high-density and outer/low-density regions, respectively. The LFs for both environments follow well the two-component function, with clear dips around $M_r = -18${\magi}, which is more prominent for the galaxies in the outer/low-density regions. Moreover, the number ratios of bright to faint components ($\Phi_g$/$\Phi^*$) measured from the two-component function of the inner/high-density regions \ylee{($\Phi_g$/$\Phi^* = 0.14$/$0.17$)} are larger than those of the outer/low-density regions \ylee{($\Phi_g$/$\Phi^* = 0.09$/$0.07$)}. All of these LF variations, depending on the environmental parameters, are common for many clusters \citep[e.g.,][]{Got05,Mer03,Mer06,Pop06,Ban10}. The best-fit parameters of the single Schechter function and the two-component function are listed in Table~\ref{table1}.

\begin{table*}
\caption{Best-fit Parameters of Luminosity Functions in Different Environments}
\label{table1} 
\centering
\resizebox{15cm}{!}{\renewcommand{\arraystretch}{1.6}
\begin{tabular}{ c c c c c c c c c }
\hline
\multicolumn{1}{c}{Region} & Function & $\Phi_g$ & $\mu_g$ & $\sigma_g$ & $\Phi^*$ & $M^*$ & $\alpha$ & $\chi^2_{\nu}$ \\
\cline{1-9}
Inner & S & - & - & - & 8.1$\pm$2.8 & 13.62$\pm$0.54 & -1.34$\pm$0.03 & 2.24 \\
Inner & G+S & 31.2$\pm$3.2 & 17.73$\pm$0.42 & 1.77$\pm$0.28 & 217.5$\pm$60.9 & 20.57$\pm$0.48 & -0.49$\pm$0.40 & 0.55 \\
Outer & S & - & - & - & 0.0$\pm$27.1 & 12.32$\pm$14907.50 & -1.57$\pm$0.04 & 7.02 \\
Outer & G+S & 10.3$\pm$1.5 & 16.67$\pm$0.28 & 1.20$\pm$0.20 & 114.3$\pm$31.0 & 19.87$\pm$0.33 & -0.75$\pm$0.20 & 2.13 \\
High-density & S & - & - & - & 4.3$\pm$1.2 & 13.90$\pm$0.44 & -1.28$\pm$0.03 & 2.12 \\
High-density & G+S & 12.3$\pm$1.2 & 17.55$\pm$0.38 & 1.68$\pm$0.26 & 71.2$\pm$22.1 & 20.34$\pm$0.54 & -0.53$\pm$0.38 & 0.41 \\
Low-density & S & - & - & - & 0.0$\pm$10.4 & 12.53$\pm$45885.54 & -1.63$\pm$0.04 & 6.57 \\
Low-density & G+S & 5.3$\pm$0.8 & 16.67$\pm$0.35 & 1.32$\pm$0.24 & 72.8$\pm$22.5 & 19.94$\pm$0.36 & -0.80$\pm$0.22 & 1.22 \\
\hline
\multicolumn{9}{p{.8\textwidth}}{note. The G and S indicate the Gaussian and Schechter function, respectively.}
\end{tabular}
}
\end{table*}

\begin{figure*}
\epsscale{1.0}
\plotone{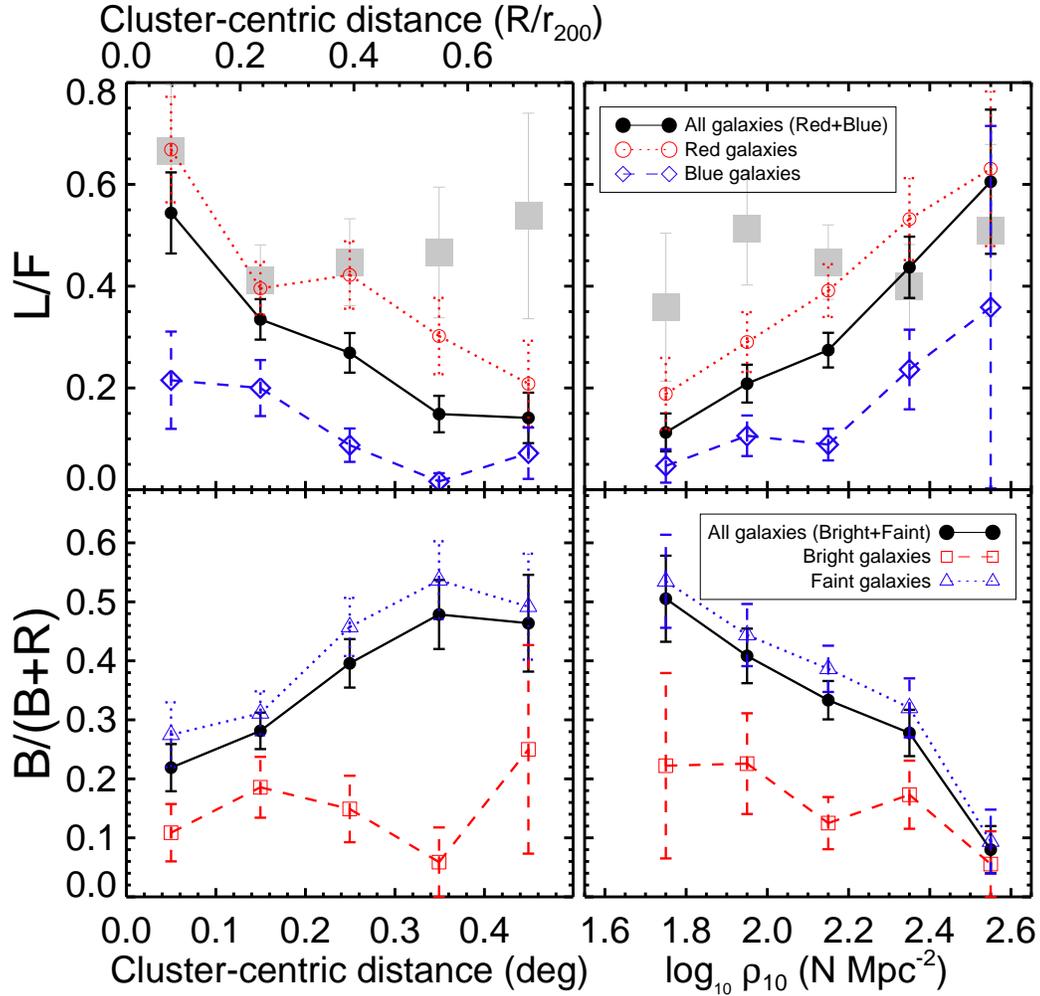}
\caption{
(Top) The luminous ($M_r < -18${\magi}) to faint ($-18${\magi} $< M_r < -14.3${\magi}) (L/F) ratio of galaxies vs. the cluster-centric distance (left panel) and the projected local density (right panel). The black solid lines with the black filled circles represent all galaxies. The red dotted lines with the red open circles and the blue dashed lines with the blue open diamonds show the red and blue galaxies, respectively. The gray squares are the L/F ratios calculated based on the definition of bright ($M_r < -19.3${\magi}) and faint ($-19.3${\magi} $< M_r < -16.5${\magi}) galaxies by \citet{Tia12}. (Bottom) The blue to total galaxy (B/(B+R)) ratio, where B and R denote the number of blue and red galaxies, respectively, plotted as a function of the cluster-centric distance (left panel) and the projected local density (right panel). The black solid lines with the black filled circles, the red dashed lines with the red open squares, and the blue dotted lines with the blue open triangles show the data for all galaxies, luminous galaxies, and faint galaxies, respectively. All error bars represent Poisson statistics.
}
\label{BFRatio}
\end{figure*}

We also estimated the number ratio of luminous to faint galaxies (L/F ratio), which roughly reflects the LF slope. The L/F ratio is a common indicator for investigating the luminosity segregation of galaxies in a cluster \citep{And02,Mer03,Ban10}. To calculate the L/F ratio, we defined ``luminous" and ``faint" galaxies as those with $M_r < -18${\magi} ($r < 18.3${\magi}) and with $-18${\magi} $< M_r < -14.3${\magi} ($18.3${\magi} $< r < 22.0${\magi}), respectively. $M_r = -18${\magi} corresponds to the magnitude at which the LF exhibits a distinct dip. Figure~\ref{BFRatio} shows the L/F ratio (top panels) for all (black filled circles), red (red open circles), and blue (blue open diamonds) galaxies, as a function of the cluster-centric distance (left panel) and the projected local density (right panel). As expected, in all cases, a strong dependence of the L/F ratio on the cluster-centric distance and local galaxy density is evident; the L/F ratio decreases with increasing the cluster-centric distance and decreasing the local galaxy density, i.e., luminous galaxies are concentrated toward the cluster center and prefer high-density regions. The L/F ratio decreases by a factor of four as we move from the center ($<$0.1 deg) to the outer regions ($>$0.4 deg). This indicates that there is a significant luminosity segregation of galaxies in the Abell 119 cluster. We also observe that the trends of the L/F ratio for red and blue galaxies (the definition is provided below) are offset (i.e., by $\sim$20-30\%). For a given environment, the overall fraction of luminous galaxies in the red sample is higher than that in the blue sample.

By using Beijing-Arizona-Taiwan-Connecticut (BATC) multicolor optical photometry, \citet{Tia12} investigated the faint-to-bright galaxy ratio (FBR) of the Abell 119 cluster, down to the limiting magnitude of $i_{BATC} = 19.5${\magi}, which is comparable to $r = 19.8${\magi} ($M_r = -16.5${\magi}) \citep{Zho03,Lup05}. In contrast to our results, they found no significant dependence of the FBR on the cluster-centric distance and local galaxy density. Following their definition of bright ($M_R < -19.5${\magi} corresponding to $M_r < -19.3${\magi}) and faint ($-19.5${\magi} $< M_R < -16.7${\magi} corresponding to $-19.3${\magi} $< M_r < -16.5${\magi}) galaxies, we also examined the variation in the L/F ratio (gray filled squares in the top panels of Fig.~\ref{BFRatio}). Consistent with the conclusions of \citet{Tia12}, we found no prominent variation in the L/F ratio and confirmed the rather flat distribution within error bars. The disagreement between the results of our analysis and that of \citet{Tia12} is probably owing to the difference between the assumed magnitude limit, discriminating between bright and faint galaxies. The magnitude limit ($M_r = -19.3${\magi}) adopted by \citet{Tia12} was $\sim$1 magnitude brighter than ours, preventing an accurate determination of the faint end of the LF. Furthermore, the shallow photometry of \citet{Tia12} might also be responsible for uncertainties in the completeness of their faint galaxy sample.

The blue-to-total galaxy ratio, B/(B+R), where B and R are the number of blue and red galaxies, respectively, indicates the evolutionary status of the cluster galaxies as a function of the environment \citep{Li09}. The blue and red galaxies denote those with $g-r$ colors bluer and redder than the red sequence $-3\sigma_{g-r}$ (see Section 2.3.1), respectively. In the bottom panels of Fig.~\ref{BFRatio}, we show the B/(B+R) ratio as a function of the cluster-centric distance (left panel) and the projected local density (right panel) for all, bright, and faint galaxies. The B/(B+R) ratio for all galaxies (black solid line with the black filled circles) increases with the cluster-centric distance and decreases with the local galaxy density. This is in accordance with the well-known observational trend between morphology (or color) and density, in which red, early-type galaxies favor the central regions of a cluster with high densities, while blue, late-type galaxies are more common on the cluster outskirts with low densities \citep[e.g.,][]{Dre80,Coo07,Bam09}.

We also examined the relative contribution of the bright and faint galaxy populations to the B/(B+R) ratio. The B/(B+R) ratio of bright galaxies (red dashed line with the red open squares) is relatively low and does not exhibit significant environmental variations within the errors, implying that the majority (i.e., $\sim$80-90\%) of bright galaxies are red, passive galaxies regardless of the environments in the Abell 119 cluster. On the other hand, it is interesting to note that the B/(B+R) ratio of faint galaxies (blue dotted line with the blue open triangles) is more sensitive to the environment, exhibiting a steeper trend than the bright galaxies. While red, faint galaxies dominate in the central, high-density regions, about half of all faint galaxies in the outer, low-density regions are blue \citep[see also][]{Pop06}. Consequently, while red galaxies dominate for both bright and faint galaxies, in the highest density regions, the fraction of blue galaxies depends strongly on luminosity in the lower density regions \citep[see also][]{Hai07,Bam09}.

\begin{figure}
\epsscale{0.9}
\plotone{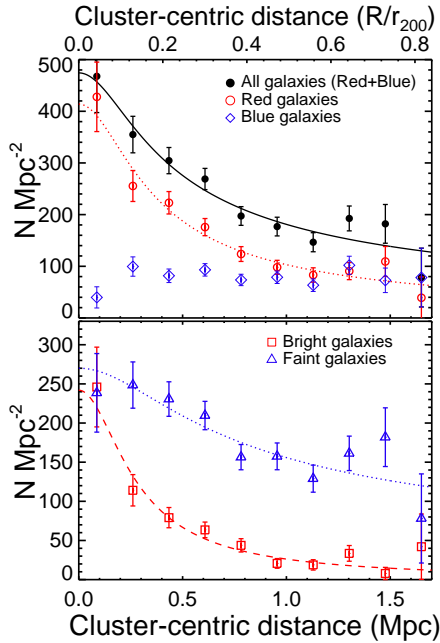}
\caption{
(Top) The projected radial density distribution of galaxies. The black filled circles show the data for all galaxies. The red open circles and the blue open diamonds show the data for red and blue galaxies, respectively. The black solid and red dotted lines are best-fit King models for all galaxies and for red galaxies, respectively. (Bottom) The projected radial density distribution of bright (red open squares) and faint (blue open triangles) galaxies. The red dashed and the blue dotted lines are best-fit King models for bright and faint galaxies, respectively.
}
\label{NumDen}
\end{figure}

In order to more specifically confirm the overall trends for the L/F and B/(B+R) ratios, we have derived the variations in the projected radial density profiles of galaxies. For that, we measured the azimuth averaged surface number density of the galaxies and fitted it with the projected King model using Equation (6). The fitting results for subpopulations with different color and luminosity are summarized in Table~\ref{table2}. In the top panel of Figure~\ref{NumDen}, the density profiles of all (black filled circles) and red (red open circles) galaxies are in good agreement with the King model profile (black solid and red dotted lines for all galaxies and for red galaxies, respectively). On the other hand, blue galaxies (blue open diamonds) exhibit an almost flat distribution without significant variation in the galaxy number density with the cluster-centric distance \citep[e.g.,][]{Hwa08,Biv09}. From the density profiles of red and blue galaxies, we again confirm that the trend of the B/(B+R) ratio for all galaxies as a function of the cluster-centric distance and local density (see the bottom panels of Fig.~\ref{BFRatio}) stems from the steep increase in the number of red galaxies toward the cluster center.

Both the density profiles for the bright (red open squares) and faint (blue open triangles) galaxies are well fitted by the King model (bottom panel of Fig.~\ref{NumDen}). It is worth noting that the bright galaxies are more centrally concentrated with \ylee{$r_c = 0.23$} Mpc and \ylee{$\sigma_0 = 241${\NMpct}}, while the faint galaxies become sparser with the cluster-centric distance and are more smoothly distributed (i.e., $r_c = 0.42$ Mpc and \ylee{$\sigma_0 = 269${\NMpct}}). This is consistent with previous results obtained for other clusters, in which fainter galaxies have more spatially extended distributions \citep{Ada98b,And02,Pra04,Par09}. Consequently, the steep increase in the number of luminous galaxies toward the cluster center, together with a wider distribution of faint galaxies, is responsible for the trend in the L/F ratio in Fig.~\ref{BFRatio}.

\begin{table*}
\caption{Best-fit King Model Profile Parameters of Radial Galaxy Density Distribution}
\label{table2} 
\centering
\resizebox{15cm}{!}{\renewcommand{\arraystretch}{1.6}
\begin{tabular}{ c c c c c }
\hline
\multicolumn{1}{c}{} & $\sigma_0$ ($N$ Mpc$^{-2}$) & $r_c$ (Mpc) & $\beta$ & $\chi^2_{\nu}$ \\
\cline{1-5}
All galaxies & 474$\pm$77 & 0.25$\pm$0.14 & 0.34$\pm$0.09 & 1.16 \\
Red galaxies & 413$\pm$77 & 0.24$\pm$0.11 & 0.48$\pm$0.11 & 0.90 \\
Bright galaxies & 241$\pm$65 & 0.23$\pm$0.10 & 0.75$\pm$0.17 & 1.15 \\
Faint galaxies & 269$\pm$41 & 0.42$\pm$0.19 & 0.29$\pm$0.17 & 1.05 \\
\hline
\multicolumn{5}{p{.5\textwidth}}{note. The $\chi_{\nu}^{2}$ is a reduced chi-square value of the King model profile fitting.}
\end{tabular}
}
\end{table*}

\subsection{Spatial Distribution of Galaxies and Luminosity Function in Overdense Regions}

Simulations and observations of the large-scale structure of the Universe have revealed that the clusters of galaxies are connected through filamentary networks \citep{Wes95,Fal05,Col05,Por07,Ber09}. Indeed, the build-up of galaxy clusters is characterized by the accretion of field galaxies or groups of galaxies through filamentary structures \citep[e.g.,][]{Ada05,Col05}. Furthermore, there is a connection between the distribution of substructures in the galaxy clusters and the larger scale cluster distributions \citep[e.g.,][]{Wes95}. Considering the possible existence of sub-clusters in the Abell 119 cluster \citep[and see below for details]{Ram07,Tia12}, it is of interest to examine large-scale structures around the Abell 119 cluster and to determine the possible filamentary structure associated with the cluster.

\begin{figure*}
\centering
\epsscale{0.8}
\plotone{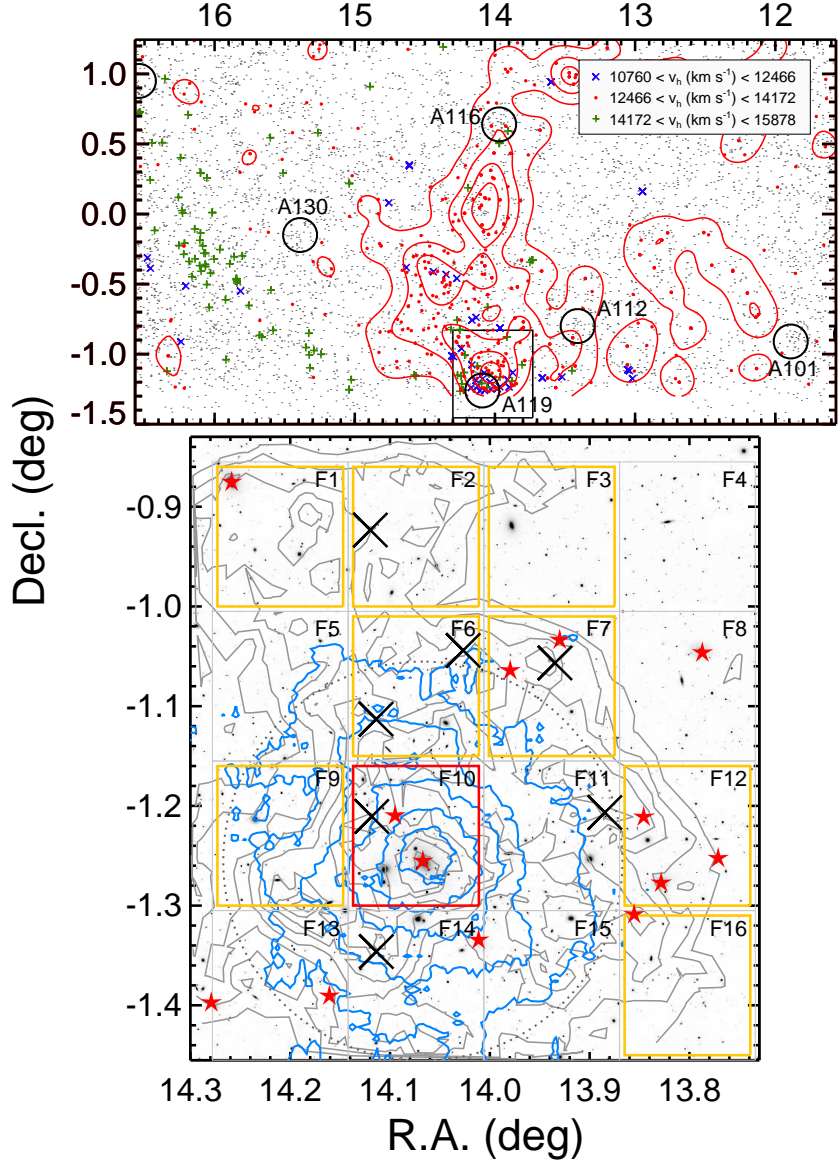}
\caption{\linespread{1.0}\selectfont{}
 { \scriptsize (Top) The large-scale distribution of galaxies around the Abell 119 cluster, obtained by using the SDSS DR9 spectroscopic data. The galaxies in the different velocity ranges are denoted by different symbols: $v_{sys} - 3\sigma_{v} < v_{h} < v_{sys} - 1\sigma_{v}$ (blue crosses), $v_{sys} - 1\sigma_{v} < v_{h} < v_{sys} + 1\sigma_{v}$ (red dots), and $v_{sys} + 1\sigma_{v} < v_h < v_{sys} + 3\sigma_{v}$ (green pluses). The gray dots show the data for the galaxies with $v_{h} < v_{sys} - 3\sigma_{v}$ or $v_{h} > v_{sys} + 3\sigma_{v}$. The red contours represent the density distribution of the galaxies with $v_{sys} - 1\sigma_{v} < v_{h} < v_{sys} + 1\sigma_{v}$, smoothed by using a Gaussian kernel of 0.1 deg. Abell clusters located in the region are denoted by large open circles with their names indicating the clusters. The large rectangle marks our observed field of view. \ylee{The scale of 0.5 deg corresponds to 1.59 Mpc at the distance of the Abell 119 cluster (i.e., 182 Mpc).} (Bottom) The spatial distribution of the galaxies in the $r$-band image of our observed field. The gray contours show the logarithmic local galaxy density of the cluster members. The large dotted circle indicates a 0.2 deg radius from the cluster center. The thin gray grid with labels defines the 16 sub-regions with the same area of 73.4 arcmin$^2$ . The central cluster region and the overdense regions are highlighted with red and yellow boxes, respectively. The large crosses denote the locations of sub-clusters defined in the previous studies \citep{Fab93,Kri97,Ram07,Tia12}. The red stars are the post-merger early-type galaxies identified by \citet{She12}. The \ylee{0.1-2.4} keV X-ray surface brightness distribution from the ROSAT observations is overlaid on the blue contours.
}
}
\label{SpacialGal}
\end{figure*}

In the top panel of Figure~\ref{SpacialGal}, we show the large-scale projected distribution of galaxies around the Abell 119 cluster, by using the SDSS DR9 spectroscopic data of galaxies with available radial velocities. We selected the galaxies in the velocity range of $v_{sys}\pm1\sigma_{v}$ (i.e. $12\,466\,$km$\,$s$^{-1} < v_{h} < 14\,172\,$km$\,$s$^{-1}$, red dots), which is comparable to the velocity range for our bright member galaxies, based on the estimated systemic velocity and velocity dispersion of the Abell 119 cluster, as explained in Section 2.3. The density distribution of these galaxies was also superimposed onto contours (see red contours), which could then be considered as a possible large-scale structure closely related to the Abell 119 cluster. For comparison, we also show the galaxies with radial velocities deviating from the systemic velocity of the Abell 119 cluster: $v_{sys} - 3\sigma_{v} < v_{h} < v_{sys} - 1\sigma_{v}$ (blue crosses) and $v_{sys} + 1\sigma_{v} < v_{h} < v_{sys} + 3\sigma_{v}$ (green pluses), which appear to constitute foreground/background large-scale structure around the Abell 119 cluster. Due to the spatial coverage limit of the SDSS observations, galaxy distribution in the south direction of the cluster center was not examined.

A remarkable feature is that a large fraction of galaxies that have similar radial velocities as the bright member galaxies in the Abell 119 cluster is distributed along the northeast direction from the Abell 119 cluster. Moreover, the distribution of the galaxies extends towards the Abell 116 cluster and the galaxies appear to form a prominent filamentary structure, connecting both clusters on the scale of $\sim$2.5 deg (i.e., 7.94 Mpc). The large-scale galaxy distribution suggests that the Abell 119 cluster is embedded in the surrounding filament of galaxies and clusters. Considering the statistical studies of large-scale intercluster filaments \citep{Bon96,Col05,Tem14}, it is tempting to speculate that the galaxies might have been falling into the Abell 119 cluster along the surrounding structure.

The bottom panel of Fig.~\ref{SpacialGal} shows the spatial distribution of galaxies within our observed field of view. We also present the contour map of surface density of the cluster member galaxies (gray contours). While the member galaxies are mainly concentrated in the central region within the radius of 0.2 deg (i.e., 0.64 Mpc, see the large dotted circle) from the cluster center, it is remarkable that the overall galaxy distribution is somewhat elongated in the northeast direction rather than exhibiting an isotropic distribution. In addition, we also present the \ylee{0.1-2.4} keV X-ray surface brightness distribution from the ROSAT observations (blue contours) \ylee{\footnote{\ylee{\url{http://www.xray.mpe.mpg.de/cgi-bin/rosat/rosat-survey}}}}. Note that the X-ray emission is only concentrated in the central region of the cluster.

In order to further examine the details of the Abell 119 cluster structure, we divided our observed field into 16 sub-regions (gray grid in the bottom panel of Fig.~\ref{SpacialGal}). Based on the definition in Section 3.2 (see also Fig.~\ref{RadialDensity}), we identified eight overdense regions (F1, F2, F3, F6, F7, F9, F12, and F16; highlighted with yellow boxes). Previous studies suggested that the Abell 119 cluster exhibits evidence of substructures in the projected galaxy distribution and the X-ray surface brightness profile \citep{Fab93,Kri97,Ram07,Tia12}. Most of these sub-clusters (large crosses) are located in and around these overdense regions.

It is interesting to note that the overdense regions are elongated along the southwest to northeast direction, following the orientation of the large-scale filament structure around the Abell 119 cluster, as shown in the upper panel of Fig.~\ref{SpacialGal}. Following \citet{Wes95}, we compared the orientation of the overdense regions with that of the larger-scale surrounding matter distribution. The projected position angle of the overdense regions was determined by the least-square fit to their locations. The position angles of the two neighbor clusters (Abell 130 and Abell 116) from the center of the Abell 119 cluster were also defined. For each neighbor cluster, we estimated the acute angle, defined as the difference between the two position angles, which is a measure of the alignment tendency between the overdense regions of the Abell 119 cluster and the neighboring cluster. The acute angles were $\sim$9 deg and $\sim$45 deg for the Abell 130 and Abell 116 clusters, respectively. Our results are in good agreement with those of \citet{Wes95}, who reported that the acute angles for 93 clusters are strongly skewed toward small values ($<$45 deg), indicating that sub-cluster pairs tend to have the same orientation as the surrounding large-scale cluster distribution. This finding suggests that anisotropic cluster formation continues through the merging of sub-clusters along large-scale filamentary features in the matter distribution \citep{Roe97,Col99,Wes95,Lee14,Paz11}.

In addition, 13 bright ($M_r < -20${\magi}) post-merger early-type galaxies (red filled stars in the bottom panel of Fig.~\ref{SpacialGal}), identified by \citet{She12}, are shown as well. About half of them ($\sim$54\%) are found in our overdense regions, and the rest are located in and around the central region. This suggests that mergers of galaxies are more frequent in overdense regions, compared with other regions such as the cluster outskirts \nylee{\citep{Cap11,Bos14}}. \citet{She12} proposed that most of these post-merger early-type galaxies were, however, formed from galaxy mergers in less-dense group environments and then entered into the dense cluster region via dark matter halo mergers. Assuming that our overdense regions are comparable to galaxy group environments, the high frequency of post-merger galaxies in overdense regions is in good agreement with the picture that galaxy groups are a suitable environment for dynamical encounters and mergers between galaxies \citep[e.g.,][]{Got01,Pip14}. Furthermore, galaxy merging can be accelerated by the perturbative effects of a collision between a cluster and a group in the course of accreting a group into a cluster \citep{Bek99,Dub99,Gne99}.

\begin{figure}
\centering
\epsscale{1.0}
\plotone{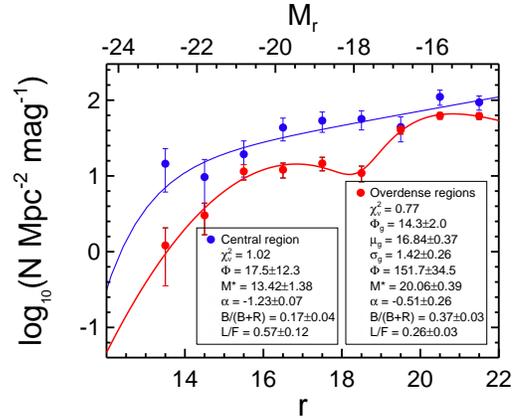}
\caption{
The composite LF of all galaxies included in the eight overdense regions (red filled circles) compared with that of the central region (blue filled circles). The error bars for all magnitude bins represent the Poisson statistics. The red and blue solid curves are the fits of a two-component function for the overdense regions and a single Schechter function for the central region, respectively. The best-fit parameters of the two-component function and the single Schechter function are shown in the boxes. The errors for all best-fit parameters were estimated from the covariance matrix. The B/(B+R) and L/F ratios and their Poisson errors are also shown.
}
\label{RegionLF}
\end{figure}

In Figure~\ref{RegionLF}, we present the composite LF of all galaxies included in the eight overdense regions (red filled circles), compared with the LF of the central region (F10, blue filled circles). The LF of the central region is satisfactorily explained by using a single Schechter function (blue solid line). By contrast, the LF of the overdense regions can only be adequately fitted by a two-component function (red solid line). A dip at $M_r\sim-18${\magi} in the LF of the overdense regions is prominent, while no such feature is found in the central region. Our result seems to be in good agreement with previous results, reporting that two-component LFs with dips are found in the intermediate- and low-density regions of clusters, while the high-density region is well represented by a single Schechter function \citep[e.g.,][]{Mer06,Pop06,Ban10}.

We also estimated the L/F ratios and the B/(B+R) ratios of the central and overdense regions; the L/F ratios are \ylee{$0.57\pm0.12$ and $0.26\pm0.03$} and the B/(B+R) ratios are $0.17\pm0.04$ and $0.37\pm0.03$ for the central and overdense regions, respectively. These values appear to be consistent with the global trend found in the Abell 119 cluster, in which the L/F ratio decreases and the B/(B+R) ratio increases with increasing the cluster-centric distance (see Section 3.2). However, the apparent effects of cluster-centric distance and local galaxy density on the L/F and B/(B+R) ratios are degenerate (see Fig.~\ref{BFRatio}). In order to evaluate the effect of enhancement of the local galaxy density (i.e., local environment) on the L/F and B/(B+R) ratios by breaking this degeneracy, we compared two overdense regions (F1 and F2) with two underdense ones (F4 and F8) at similar cluster-centric distances of $\sim$0.3-0.4 deg from the cluster center (i.e., similar global cluster environment). The L/F and B/(B+R) ratios slightly differed for these overdense and underdense regions: the L/F ratios are \ylee{$0.16\pm0.04$ and $0.06\pm0.03$} and the B/(B+R) ratios are $0.37\pm0.07$ and $0.56\pm0.12$, for the overdense and underdense regions, respectively. This result suggests that the cluster galaxies located in the locally overdense regions reach a higher fraction of luminous, red galaxies, although they are located at similar cluster-centric distance as the other galaxies in the outer underdense regions. In the context of ``preprocessing" \citep{Bla04,Fuj04,Tan05,Coo07,McI08}, it is understandable that locally higher galaxy density regions, such as galaxy groups and substructures, are likely to be already in a more advanced evolutionary stage, in which the galaxies have completed their star formation and become more massive, before falling into the cluster \citep{Li09,DeL12}.

\section{Discussion and Summary}

We have presented a detailed analysis of the LF for galaxies in the Abell 119 cluster using photometric data taken with the MOSAIC II CCD mounted on the Blanco 4m telescope at the CTIO. Accurate membership assignment of cluster galaxies has been achieved by considering the galaxy radial velocities as well as the color-magnitude relation for bright galaxies and the scaling relation for faint galaxies. The main results of our study are the following.

\begin{enumerate}[(i)]
\item The LF of all galaxies in the observed region was obtained in the r-band for $13${\magi} $< r < 22${\magi} ($-23.3${\magi} $< M_r < -14.3${\magi}). The LF exhibits a clear dip at $r \sim 18.5${\magi} ($M_r \sim -17.8${\magi}), and the sum of a Gaussian and a Schechter function is a more suitable representation of the data compared with a single Schechter function. The general shape of our LF is consistent with that reported in the previous study by \citet{Pra05}.

\item To investigate the LF as a function of the environment within the Abell 119 cluster, we compared the LFs for two different regions characterized by inner, high-density and outer, low-density galaxies. The overall LF in the outer, low-density regions is systematically steeper than that in the inner, high-density region. The dip around $M_r = -18${\magi} in the outer, low-density regions is more prominent than that in the inner, high-density region.

\item The L/F ratio decreases with increasing the cluster-centric distance and decreasing the local galaxy density, indicating a significant luminosity segregation of galaxies in the Abell 119 cluster. The B/(B+R) ratio for all galaxies increases with increasing the cluster-centric distance and decreases with increasing the local galaxy density, which is consistent with the morphology-density and color-density relations. For the faint galaxies, the B/(B+R) ratio is more sensitive to different environments, exhibiting a steeper trend compared with that for the bright galaxies.

\item From the variation in projected radial number density profiles of galaxies, we confirmed the overall trends of the L/F and B/(B+R) ratios as a function of the cluster-centric distance. The number density profile for red galaxies is adequately described by the King model, with a steep increase of these galaxies toward the cluster center. Blue galaxies show an almost flat distribution as a function of the cluster-centric distance. In the number density profiles for the bright and faint galaxies, the bright galaxies are more centrally concentrated, while the faint galaxies are more uniformly distributed.

\item 
We defined overdense regions as the regions in which the local galaxy density is above the best-fit King model profile for a given cluster-centric distance. We found that the combined LF of overdense regions exhibits a two-component function, with a distinct dip at $M_r\sim-18${\magi}, while no such feature was observed for the central region. We also compared the L/F and B/(B+R) ratios between overdense and underdense regions at similar cluster-centric distances and found that these ratios have slightly different values. This suggests that overdense regions are undergoing a different stage of dynamical evolution compared with the underdense ones, even though both regions are located on the cluster outskirts.
\end{enumerate}

One of the most noticeable results of our analysis is that the Abell 119 cluster appears to be closely associated with a large-scale filamentary structure: the overall galaxy distribution as well as overdense regions in the Abell 119 cluster are found to be aligned with the surrounding filamentary matter distribution on larger scales (see Fig.~\ref{SpacialGal}). This strongly suggests that the Abell 119 cluster is still in the stage of cluster assembly through an anisotropic accretion of groups of galaxies along the filamentary structure \citep[e.g.,][]{Wes95}. The overdense regions are likely associated with subgroups that migrate in perpendicular to the line-of-sight from the cluster outside, forming elongated structure along the infall direction \citep[e.g.,][]{Vij15}. It is worth noting that the results of many observations have supported the possible scenario of a recent or ongoing merger between the main cluster and sub-clusters in the Abell 119 cluster \citep{Edg90,Bli98,Fer99,Vik09,Ros10}. In this respect, we may be witnessing the epoch of cluster build-up by the assembly of multiple galaxy groups in the context of hierarchical structure formation.

Mergers between galaxies are expected to be much more frequent in galaxy groups owing to the slow orbital motions of the galaxies within such relatively low-density environments compared with galaxy clusters \citep{Ost80}. In this respect, galaxy groups exhibit an interesting feature in their galaxy LF owing to the fact that galaxy mergers change the luminosity distribution of galaxies and therefore influence the LF shape. It has been shown that galaxy groups with low-velocity dispersions exhibit dips in their LFs \citep{Mil04}. A possible explanation is that dynamical friction causes more rapid galaxy merging in the center of the group, which depletes intermediate-luminosity galaxies to form a few luminous ones, resulting in the distinct dip observed in the LF \citep[e.g.,][]{Tre02,Mil04,Mil06}. Thus, as the galaxy group evolves, the bright end of the LF is strongly modified, while the faint end retains its initial shape \citep{Mil04}.

If groups survive on the outskirts of the Abell 119 cluster after they fall into the cluster, they might be seen as overdense regions in which the local galaxy density is higher than the mean value at a given cluster-centric distance. We suspect that the galaxies in these overdense regions may not yet have experienced the severe environmental effects of the cluster, preserving their dynamically cold behavior of a group environment. In this way, the overdense regions can retain their two-component LFs with a dip, which was shaped in the group environment before they are fully accreted into the cluster. The lack of X-ray emission in most of the overdense regions (see Fig.~\ref{SpacialGal}) would also imply that they are not yet virialized, suggesting that most of the galaxy mergers have taken place in recent epochs. As a result, it would be likely to observe a two-component LF with a dip in the overdense regions of the Abell 119 cluster similar to that found in group environments (see Fig.~\ref{RegionLF}). By contrast, in the high-density cluster core, high-velocity dispersions inhibit the merging of dwarf or intermediate-mass galaxies to produce luminous galaxies. In addition, the lack of a dip in the LF of the central region of the Abell 119 cluster can be also attributed to the tidal or collisional disruption of dwarf galaxies near the cluster center \citep{Mer06,Pop06}.

\begin{figure}
\centering
\epsscale{1.0}
\plotone{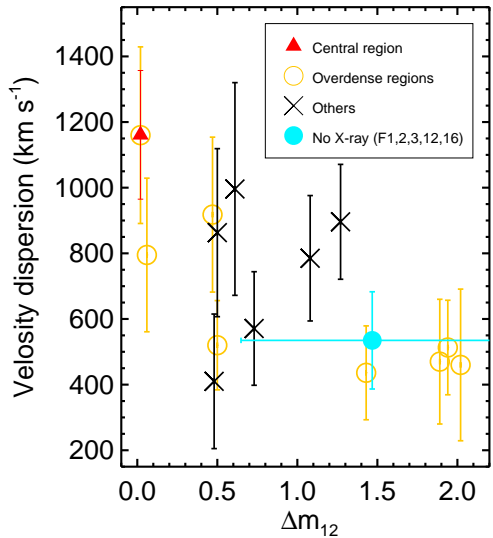}
\caption{
The velocity dispersion of our sub-regions as a function of the difference $\Delta m_{12}$ between the $r$-band magnitudes of the brightest and second brightest galaxies. Overdense regions and other sub-regions are plotted as yellow open circles and black crosses, respectively. The cyan filled circle denotes the mean value for the overdense regions with no X-ray emission, and the red filled triangle marks the central region. The error bars represent the standard deviations of the values.
}
\label{DelM12}
\end{figure}

The above explanation seems to be supported by the relation between the velocity dispersion of sub-regions and the difference between the $r$-band magnitudes of the brightest and second brightest galaxy (i.e., $\Delta m_{12}$) in each sub-region (see Figure~\ref{DelM12}). The overdense regions (yellow open circles) appear to exhibit systematically lower velocity dispersions and larger $\Delta m_{12}$ values. In particular, the overdense regions with no X-ray emission (i.e., F1, F2, F3, F12, F16, large cyan filled circle) exhibit a distinctly smaller mean velocity dispersion ($534\pm148$\kms{}) and a larger mean $\Delta m_{12}$ ($1.47\pm0.82${\magi}) compared with that of the central region (i.e., $1161\pm196$\kms{} and 0.02{\magi}). This result is consistent with the scenario in which galaxy mergings are more frequent in a system with low-velocity dispersion, leading to a large magnitude gap between brightest galaxies \citep[e.g.,][]{Mil04,Dar07,Dar10,Har12,Goz14}.

Further circumstantial support for frequent merging events in a low-velocity dispersion system comes from the direct discovery of merger relics in cluster galaxies. \citet{She12} recently reported that a surprisingly large portion (38\%) of massive early-type galaxies exhibit post-merging features in the Abell 119 cluster. Considering simple theoretical estimations of the merger timescale as a function of the relative speeds between galaxies, this is an unexpected result in a cluster environment. They suggested that a large fraction of massive early-type galaxies are formed through galaxy mergers in a low-density environment such as a galaxy group, and then fall into the cluster via cluster-group halo mergers. The post-merger features are typically visible in the current cluster for about a few Gyr \citep{She12,Yi13,Ji14}. In this respect, it is worth noting that a large fraction (54\%) of the post-merger galaxies in the Abell 119 cluster are detected in our overdense regions, supporting again the notion that overdense regions are comparable to the environment of galaxy groups where galaxy mergers are frequent. Consequently, the bimodal shape of the LFs as well as the high frequency of post-merger galaxies in the Abell 119 cluster indicate that the group environment is responsible for ``preprocessing" galaxies on smaller scales or in local structures prior to the assembly of galaxy groups into the larger global cluster scale. \ylee{More studies for other dynamically young clusters are required to confirm the universal feature of their bimodal LF and then to constrain the assembly history of galaxy clusters.}

\acknowledgments
\ylee{We thank the anonymous referee for helping us to significantly improve the original manuscript.} This research was supported by Basic Science Research Program through the National Research Foundation of Korea (NRF) funded by the Ministry of Education, Science, and Technology (2015R1A2A2A01006828). Support for this work was also provided by the NRF of Korea to the Center for Galaxy Evolution Research (No. 2010-0027910). This study was also financially supported by research fund of Chungnam National University in 2014. Y.L. was supported by the Korea Astronomy and Space Science Institute under the R\&D program supervised by the Ministry of Science, ICT and Future Planning. S.K.Y acknowledges support from the Korean National Research Foundation (NRF-2014R1A2A1A01003730). Y.K.S acknowledges support by FONDECYT \nnylee{Postdoctoral Fellowship (No. 3130470)}.

\bibliography{abell119}

\end{document}